\documentclass[twocolumn]{aastex631}

\usepackage{amsmath}
\usepackage{subfigure}
\usepackage{float}
\usepackage{gensymb}
\usepackage{placeins}
\usepackage{appendix}
\usepackage{cancel}
\usepackage{mdframed}
\usepackage{orcidlink}

\usepackage[mathlines]{lineno}

\definecolor{darkorange}{cmyk}{0,0.63,0.88,0}

\begin{document}

\title{Stellar Spectroscopy Using Diffraction Grating, CMOS Monochrome Sensor and Reflecting Telescopes}

\author[0009-0004-5627-7196]{Abhinav Roy}
\altaffiliation{{These two authors contributed equally to this paper and are designated as co-first authors.}}
\affiliation{School of Physical Sciences, National Institute of Science Education and Research, HBNI, Jatni 752050, Odisha, India}

\author[0009-0002-1114-8564]{Niti Singh}

\thanks{Correspondence: Niti Singh (nitisingh219@gmail.com)}

\affiliation{School of Physical Sciences, National Institute of Science Education and Research, HBNI, Jatni 752050, Odisha, India}

\begin{abstract}
We present the design and testing of a compact stellar spectrometer developed for undergraduate and outreach applications. The spectrometer module—comprising an inexpensive 600 lines/mm diffraction grating, a CMOS monochrome sensor, and a 3D-printed mount—can be constructed at modest cost and integrated with reflecting telescopes commonly available at educational institutions. Calibration was performed using a helium emission source in the laboratory and Vega as a spectrophotometric standard, supported by a custom Python-based data-reduction pipeline for wavelength calibration and spectral stacking. The spectrometer successfully recorded usable spectra of bright stars including Vega, Sirius, Procyon, Capella, and Betelgeuse, covering spectral types A through M. The results demonstrate that meaningful stellar spectroscopy can be achieved by institutions with existing telescope infrastructure, providing a practical framework for student-led astronomical instrumentation projects.
\end{abstract}

\section{Introduction}

Since the 19th century, stellar spectroscopy has revolutionized our understanding of astrophysics, beginning with Fraunhofer's discovery of dark absorption lines in the solar spectrum \citep{fraunhofer1814, Kirchoff1860FL, Kirchoff1860Sun}. By the early 20th century, Annie Jump Cannon's Harvard classification and Cecilia Payne's work on stellar atmospheres established spectroscopy as the foundation of stellar astrophysics, linking spectral signatures to temperature, chemical composition, and stellar evolution \citep{1925Payne, maury1897spectra}.

Today, high-resolution spectroscopy remains a cornerstone of observational astrophysics, enabling precise measurements of stellar abundances, radial velocities, and the detection of exoplanetary atmospheres \citep{2005DavidG, 2005Martin}. Recent application of HIRES (High Resolution Echelle Spectrometer) on the Keck I Telescope, employs a cross-dispersed echelle-grating spectrograph, achieving spectral resolutions between 25000 to 85000 across a wavelength range of 0.3 to 1.0 microns \citep{HIRES}. This high resolution allows for precise measurements of stellar parameters and has been instrumental in various astrophysical studies. Inspired by these sophisticated instruments, our project aims to construct a compact diffraction-grating spectrometer integrated with a reflecting telescope and a CMOS (Complementary Metal-Oxide-Semiconductor) sensor.

\begin{figure*}[p]
\centering
\includegraphics[width=\columnwidth]{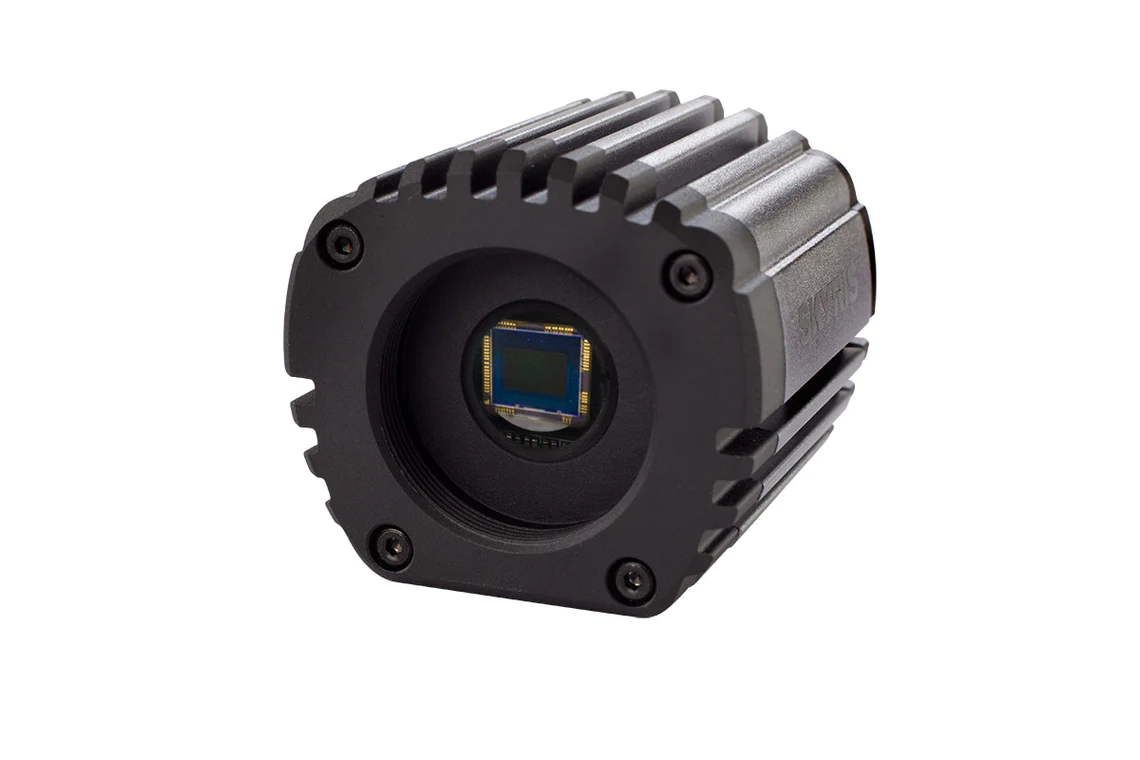}
\hfill
\includegraphics[width=\columnwidth]{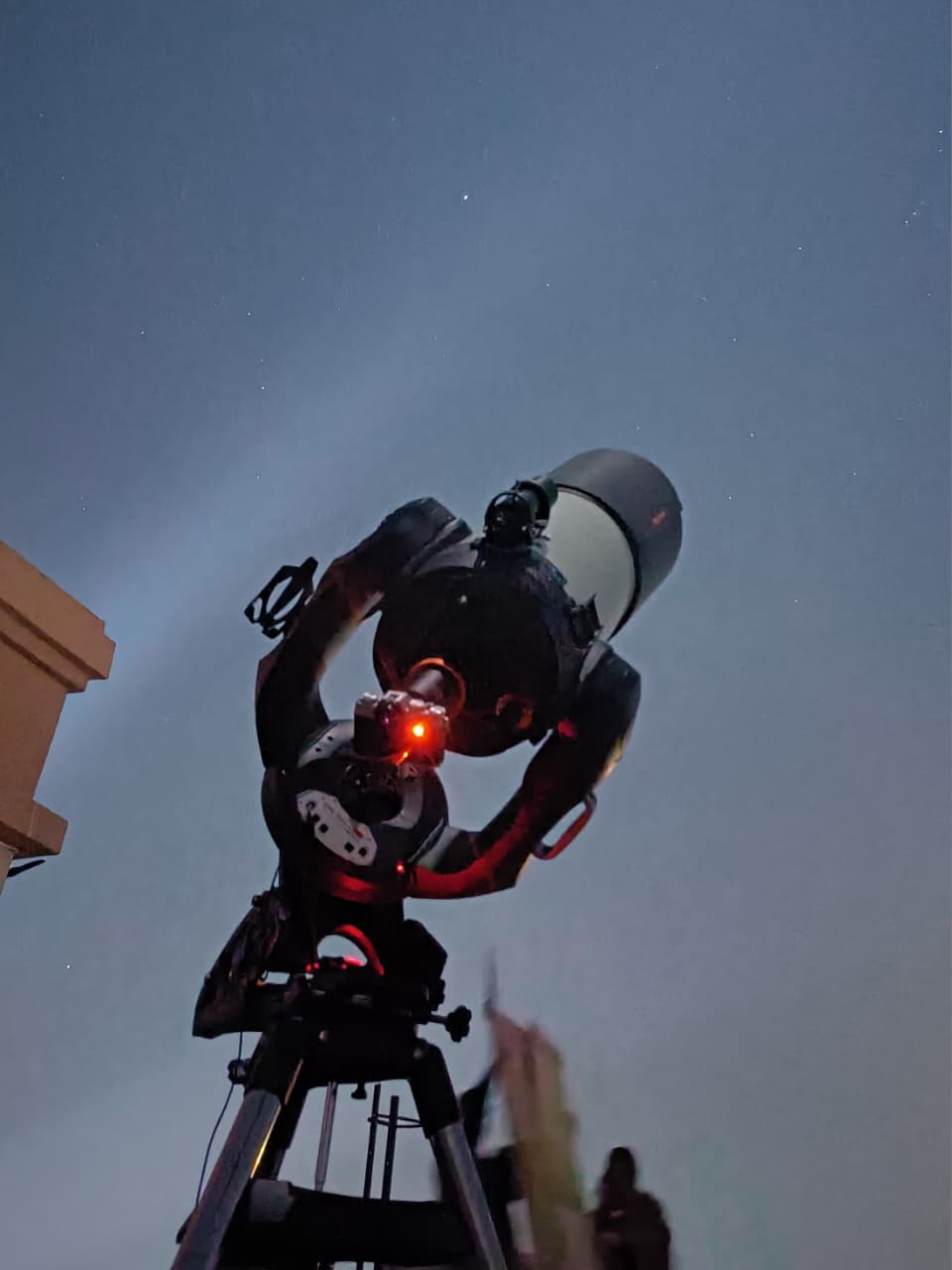}
\caption{Left: Skyris 236M CMOS monochrome sensor used for spectral imaging. Right: 11-inch Celestron CPC Deluxe 1100 HD telescope with computerized tracking, used for primary stellar observations.}
\label{fig:telescopes}
\end{figure*}

While operating at a lower resolution, the instrument embodies similar fundamental techniques, providing a hands-on platform to observe and analyze stellar spectra across multiple spectral classes. This approach not only offers practical experience in astronomical instrumentation but also contributes to the growing field of accessible astronomical spectroscopy. Using a 600 lines/mm diffraction grating with a CMOS monochrome sensor, and a custom 3D-printed mount coupled to an 11-inch Schmidt-Cassegrain telescope (Figure \ref{fig:telescopes}), we constructed a spectrometer capable of dispersing starlight into well-defined spectra. Our targets -- Sirius (A1V, the brightest star in the night sky), Vega (A0V, a standard photometric reference), Procyon (F5IV-V, a nearby subgiant), Capella (G8III + G0III, a quadruple system with a primary-pair of giants), and Betelgeuse (M2Ia, a red supergiant) \citep{Gray_2003_Sir_Vega, Perkins_BG_cap, procyon0, Torres_2015_Capella_bin} -- were chosen to span a wide range of stellar types and evolutionary stages. Each of these stars carries historical significance in spectroscopy: Vega as the flux calibration standard, Sirius as an early benchmark in line-profile studies, and Betelgeuse as one of the first stars with a resolved stellar disk.

The overarching aim was to demonstrate that stellar spectra from such diverse stars could be recorded and calibrated while being feasible at the institute or undergraduate research level, broadening access to hands-on observational experience. While the telescopes used in this work represent significant institutional investments, the spectrometer module itself—diffraction grating, sensor, and 3D-printed mount—can be assembled at modest cost, making the approach accessible to institutions that already possess suitable telescopes. In doing so, the project highlights both the potential and limitations of amateur-level spectroscopy, while offering an accessible framework for expanding its application in both research and education.

This activity is primarily designed for undergraduate physics students (second year and above), with extensions suitable for advanced high-school outreach and junior graduate-level instrumentation projects. For physics-focused cohorts, learning outcomes include wavelength calibration, spectral classification, uncertainty estimation, and comparison with reference spectra. For computationally oriented students, the development and validation of Python-based data-reduction and spectral-stacking pipelines can form a central learning objective. Engineering-oriented implementations may emphasize optical alignment, mechanical design of the spectrometer mount, and performance optimization of slit width and grating geometry.

\section{Theoretical Background}

\subsection{Stellar Spectroscopy Principles}

Every star carries its own spectral fingerprint - an intricate code of light that reveals its hidden physical and chemical identity. Through stellar spectroscopy, we decode this light to study the temperature, composition, and motion of astrophysical bodies far beyond our reach. When a star emits radiation, its photosphere acts as a continuous light source, while the cooler gases in its outer layers absorb specific wavelengths corresponding to electronic transitions of various atoms and ions. These absorbed wavelengths appear as dark absorption lines within the continuous spectrum, each marking the presence of an element such as hydrogen, calcium, or iron \citep{carroll2017introduction}. 

The systematic study of these spectral lines forms the foundation of stellar classification and atmospheric modeling. For instance, A-type stars like Vega and Sirius display striking hydrogen Balmer lines, indicative of high photospheric temperatures ($\sim$10,000 K)\citep{Gray_2003_Sir_Vega}. In contrast, G- and K-type stars such as Procyon or Capella exhibit stronger metallic lines, while M-type giants like Betelgeuse reveal molecular absorption bands from titanium oxide in the red and near-infrared regions\citep{gray2021observation, morgan1942atlas}. These differences can trace stellar temperature and ionization states to yield information about stellar evolution and surface gravity. However, terrestrial observations are affected by telluric absorption from Earth's atmosphere, which can be corrected to recover intrinsic stellar signature \citep{thorne1999spectrophysics}. By studying these features, even small-scale spectroscopic experiments can capture the essence of stellar physics that drives modern astrophysical research.

\subsection{Diffraction Grating Theory}

For this project, we use a diffraction grating, which is an optical grating that diffracts light into its constituent wavelengths into different directions. Unlike a prism, which relies on refraction and introduces chromatic aberrations, a grating uses the principle of interference between light waves diffracted by a regular array of slits or grooves. This allows for sharper, more evenly spaced spectral lines and enhanced wavelength resolution, an essential advantage for observing subtle stellar absorption features \citep{hecht2023optik}. The choice of a grating-based spectrometer thus provides both affordability and scientific accuracy, making it particularly well-suited for educational and research applications in stellar spectroscopy.

A transmission diffraction grating disperses incident light according to the grating equation:
\begin{equation}
d \sin\theta = n\lambda
\label{eq:grating}
\end{equation}
where $d = 1/N$ is the grating spacing (N = 600 lines/mm for our grating), $\theta$ is the diffraction angle, $n$ is the diffraction order, and $\lambda$ is the wavelength \citep{tonkin2013practical}. For optimal spectral resolution with minimal overlapping orders, we operated in first-order diffraction ($n=1$), yielding angular dispersion of approximately $12.1\degree$ to $24.8\degree$ across the visible range (350-700 nm).


\begin{figure*}[p]
\centering
\includegraphics[width=\columnwidth, height=6.9cm]{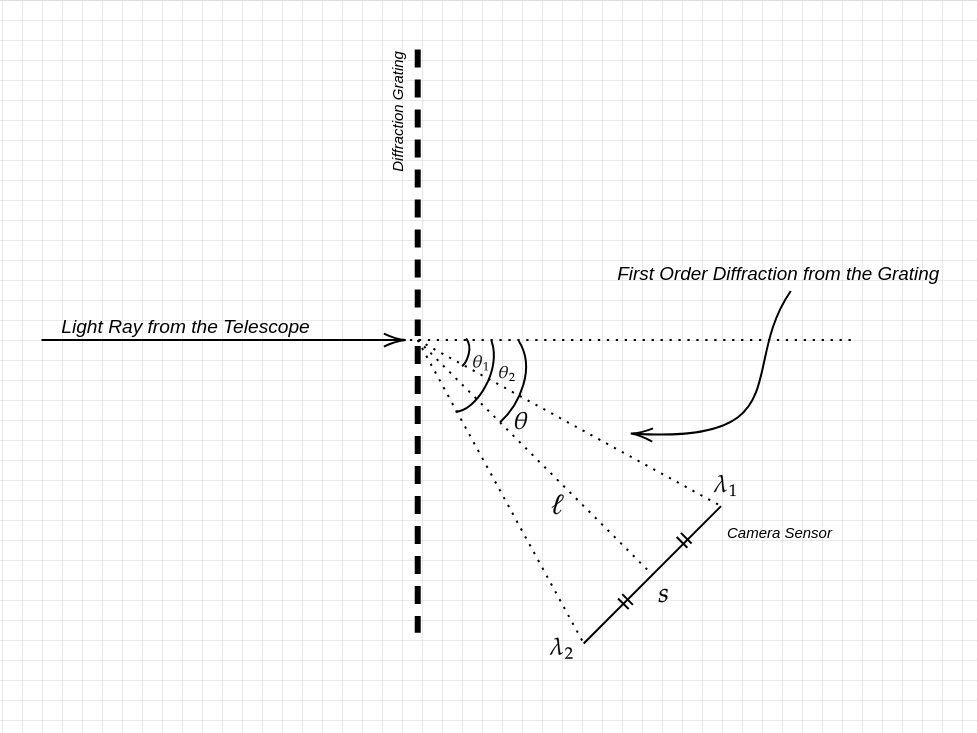}
\hfill
\includegraphics[width=\columnwidth, height=6.9cm]{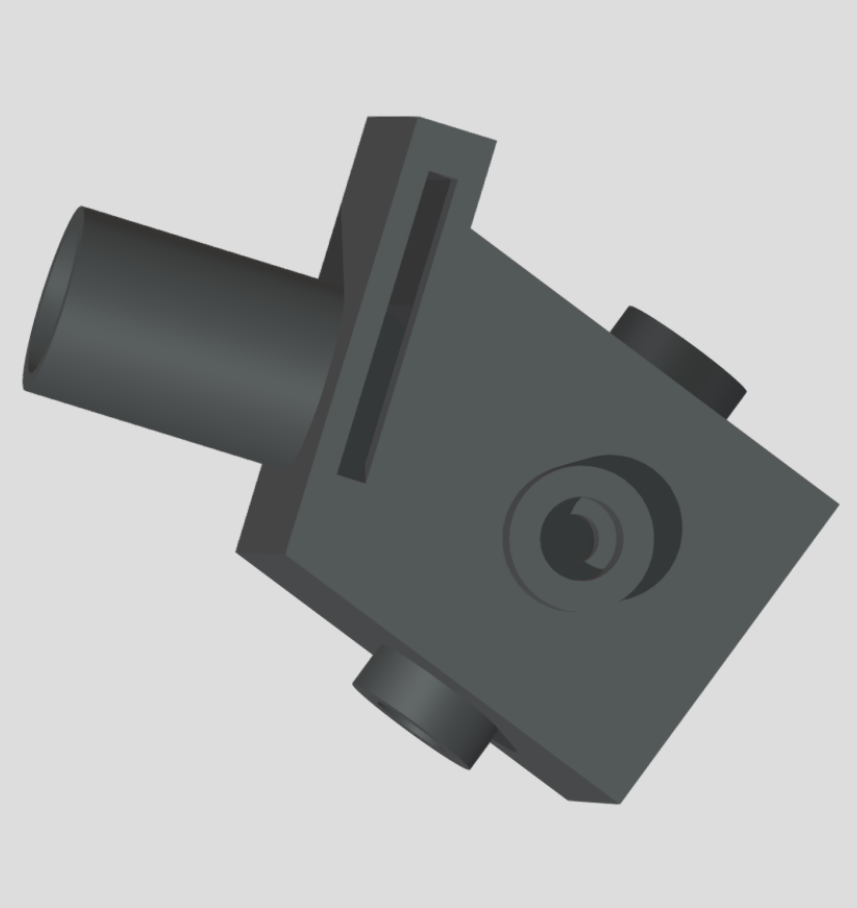}
\caption{Left: Ideal geometric alignment showing the diffraction grating, sensor position, and angular geometry for first-order diffraction. The distance $l$ between grating and sensor was optimized to capture the full visible spectrum. Right: 3D model of the custom 3D-printed spectrometer mount designed in SolidWorks. The mount securely holds the diffraction grating and CMOS sensor while interfacing with standard telescope eyepieces.}
\label{fig:spectrometer-design}
\end{figure*}

\section{Methods}

\subsection{Instrument Design and Construction}

Our spectrometer consists of three main components: a 600 lines/mm transmission diffraction grating, a Skyris 236M CMOS monochrome sensor, and a custom 3D-printed mount that attaches to standard telescopes. The key specifications of our equipment are summarized in Table~\ref{equipment-table}.

\begin{table}[h]
\centering
\begin{tabular}{cc}
\hline
\textbf{Component} & \textbf{Specification} \\
\hline
Diffraction grating & 600 lines/mm transmission \\
CMOS sensor & Skyris 236M monochrome \\
Sensor dimensions & 5.44 mm $\times$ 3.67 mm \\
Primary telescope & 11-inch CPC Deluxe 1100 HD \\
Secondary telescope & 8-inch SkyWatcher Quattro 200P \\
Slit width range & 1.0 - 1.5 mm \\
Wavelength coverage & 350 - 700 nm \\
\hline
\end{tabular}
\caption{Key equipment specifications used in our stellar spectrometer}
\label{equipment-table}
\end{table}

The equipment configuration was largely determined by available institutional resources - the 8-inch and 11-inch telescopes and Skyris 236M CMOS (Complementary Metal-Oxide-Semiconductor) sensor were on hand, providing an opportunity to explore how amateur-grade equipment compares to professional observatories such as the Hubble Space Telescope(HST). Importantly, while the telescopes represent substantial capital equipment typically available at university observatories or well-equipped astronomy clubs, the spectrometer components themselves—transmission grating, 3D-printed mount, and imaging sensor—are relatively simple and widely available from educational lab suppliers. The 600 lines/mm transmission diffraction grating balances spectral dispersion, wavelength coverage, and physical compactness, providing sufficient resolution to capture prominent stellar absorption features while maintaining a profile mountable to standard 1.25-inch eyepiece adapters. Higher line densities (e.g., 1200 lines/mm) would improve resolution but reduce wavelength coverage for a fixed sensor size, while lower densities (e.g., 300 lines/mm) might sacrifice the ability to distinguish individual spectral features. The Skyris 236M has been discontinued, but the design and approach can be adapted to any alternative CMOS or CCD (Charge-Coupled Device) sensors with appropriate specifications (sensor size $\geq$5mm, 16-bit depth, 2-4 $\mu$m pixels) and most importantly low read noise for detecting faint absorption lines.

The design is intentionally flexible and scalable. While larger telescope apertures (8-11 inches) reduce exposure times, smaller telescopes (4-6 inches) can observe bright stars (V $<$ 2 mag) with longer integrations. Adjustments to slit width, exposure duration, and stacked frame count can accommodate various telescope sizes and observing conditions. This adaptability underscores the central aim: demonstrating that the spectrometer module can leverage existing telescope infrastructure—whether at university observatories, astronomy clubs, or well-equipped schools—to enable meaningful stellar spectroscopy through rigorous scientific methodology rather than specialized instrumentation.

The optimal sensor-grating geometry for our setup was derived from diffraction theory to capture the complete visible spectrum (350-700 nm) across the sensor width. We focused on first-order diffraction ($n=1$) because higher orders would require prohibitively large diffraction angles that exceed our compact sensor geometry, demanding a much larger sensor or impractically small grating-to-sensor distance, making it physically impossible. The desired wavelength range $\lambda_1$ to $\lambda_2$ directly yields the corresponding diffraction angles, from the diffraction geometry illustrated in Figure~\ref{fig:spectrometer-design}:
\begin{equation}
\theta_1 = \sin^{-1}(\lambda_1 N) \quad \text{and} \quad \theta_2 = \sin^{-1}(\lambda_2 N)
\end{equation}
Here, $\theta_1$ and $\theta_2$ correspond to the diffraction angles for the lower ($\lambda_1$ = 350 nm) and upper ($\lambda_2$ = 700 nm) limits of the spectrometer’s wavelength coverage, respectively.
While the required distance $l$ between grating and sensor is given by:
\begin{equation}
l = \dfrac{s}{2\tan\left(\dfrac{\theta_2 - \theta_1}{2}\right)}
\end{equation}
Here, \textit{s} denotes the full active sensor width (5.44 mm), which is symmetrically divided by the optical axis at distance \textit{l}, yielding two equal half-widths (indicated by ‘=’ on \textit{s} in Figure~\ref{fig:spectrometer-design}). With $\theta_1 = 12.1\degree$ and $\theta_2 = 24.8\degree$, we calculated $l \approx 25$ mm, positioning the spectrum centrally on the sensor.

The spectrometer housing was designed in SolidWorks and fabricated via fused deposition modeling using PLA (Polylactic Acid) plastic. The mount integrates a standard 1.25 inch telescope eyepiece adapter, an adjustable grating holder for fine angular tuning, and a rigid sensor fixture maintaining a 15 mm focal distance from the camera interface. This configuration ensures precise optical alignment and ease of integration with both the 8-inch and 11-inch telescopes used for observations. Figure~\ref{fig:spectrometer-design} illustrates the 3D model and the ideal optical configuration.

\subsection{Slit System}
\label{sec:slit-system}

During preliminary tests using a collimated helium lamp placed at large distances (hundreds of meters), we observed that spectral lines appeared broadened when viewed through the telescope. Although both stars and the lamp act as point sources at infinity, telescope optics and atmospheric seeing cause them to form finite disks. Consequently, the stellar image was spread over multiple sensor rows, producing broadened spectral lines and reducing effective resolution. To prevent this, we used a narrow entrance slit to confine the incoming light and preserve fine spectral detail.

Entrance slits are a standard component of professional astronomical spectrographs, including long-slit and multi-object instruments, where they define spatial sampling and spectral resolution. The slit-based design adopted here mirrors these professional approaches, differing primarily in scale and resolving power rather than principle. 

We implemented an adjustable precision-blade slit system with widths between 1.0 and 1.5 mm, positioned ahead of the spectrometer grating. The slit acts as a spatial filter, admitting light from a narrow region of the stellar image into the spectrometer. Selecting the optimal slit width required balancing three competing factors: spectral resolution, light throughput, and diffraction effects.

The 1.0-1.5 mm range provided the best compromise for our sensor dimensions (5.44 mm width) and telescope focal ratios (f/4 to f/10), maintaining adequate light levels for reasonable exposure times while preserving the resolution needed to resolve individual absorption features.

\subsection{Calibration and Testing}

\begin{figure}[H]
\centering
\includegraphics[width=\columnwidth]{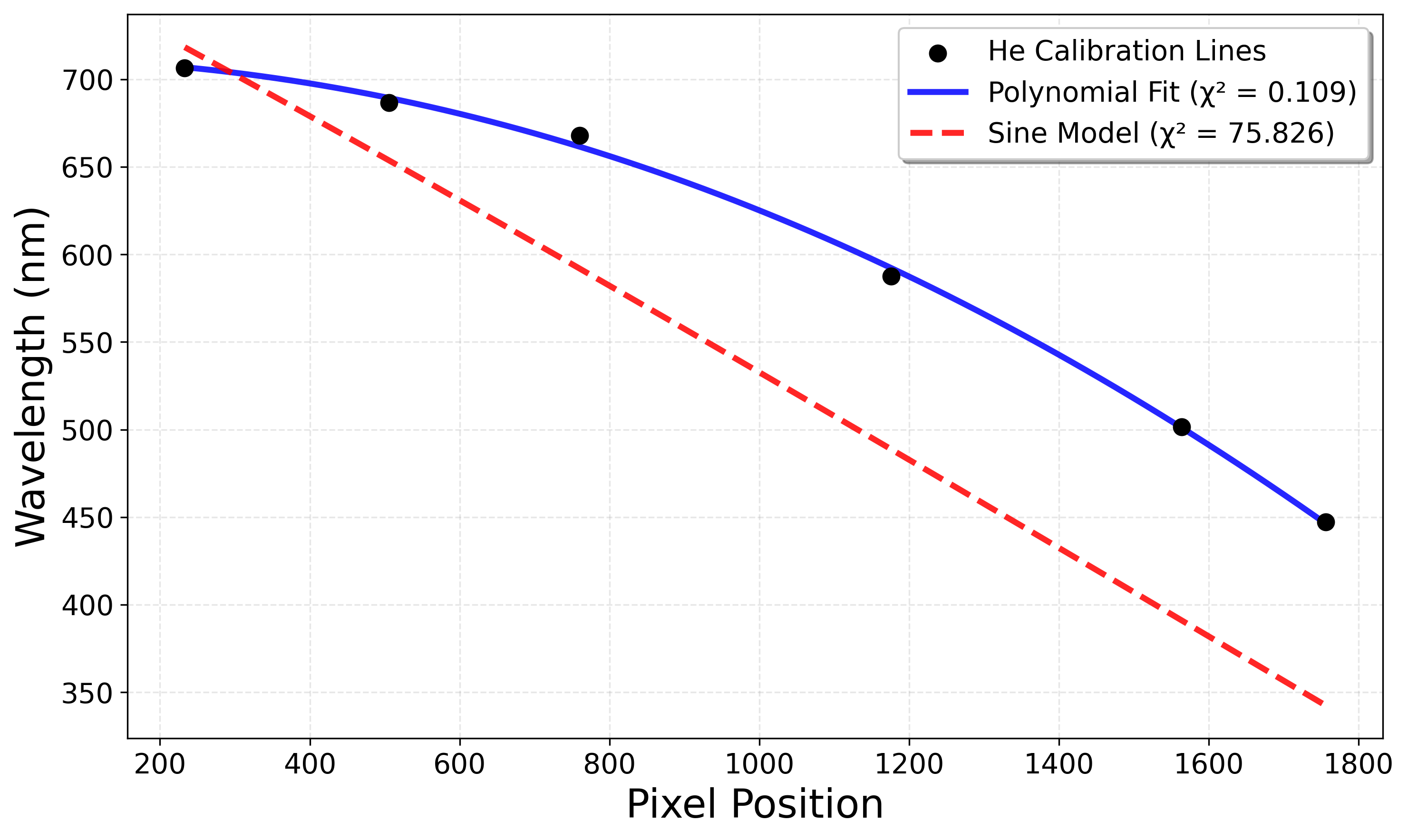}
\caption{Wavelength calibration using helium emission lines captured with the 8-inch telescope. The quadratic polynomial fit (blue solid line) follows the calibration data closely, whereas the sinusoidal relation predicted by the grating equation (red dashed line) deviates systematically. The polynomial approximation is therefore appropriate for our optical geometry, where minor misalignments and the asymmetric pixel-to-angle projection across the small sensor field justify deviating from the theoretical sinusoidal form.}
\label{fig:he-calibration}
\end{figure}

The spectrometer was first validated under controlled laboratory conditions using a helium emission lamp as wavelength standard. The helium lamp provided six prominent lines (447.1-706.5 nm) across the visible range. Initial alignment with a pinhole collimator produced internal reflections, which were mitigated by optical baffling and path optimization. Pixel positions of the identified emission peaks were matched to known wavelengths to obtain a polynomial wavelength calibration:
\begin{equation}
\lambda(\text{nm}) = -0.000079x^2 - 0.0381x + 711.502,
\end{equation}
where $x$ denotes the pixel column. Although the grating equation (Equation~\ref{eq:grating}) implies a sinusoidal relation, a polynomial fit was preferred due to minor optical misalignments and the asymmetric pixel-to-angle projection across the small sensor field (Figure~\ref{fig:he-calibration}). Figure~\ref{fig:he-calibration} compares the polynomial and sinusoidal fits: the polynomial reproduces the measured line positions across the full sensor width, whereas the sinusoidal form derived from the ideal grating geometry deviates systematically from the calibration data.

Following laboratory calibration, we tested the spectrometer on two reflecting telescopes: an 8-inch (f/4) and an 11-inch (f/10). The larger aperture increased light-gathering power and allowed shorter exposures, but resulted in broadened stellar images at the focal plane, reducing resolution. We therefore used an adjustable slit as detailed earlier, and computerized tracking on the 11-inch to focus on targets for sequential exposures and stacking.

Using helium lamp lines, we measured the instrumental resolving power. Emission line profiles exhibited a full-width-at-half-maximum (FWHM) of $\approx101\pm8$\,px, representing spectral broadening from slit width, optical aberrations, and detector sampling. The wavelength dispersion $d\lambda/dx$ was computed by differentiating the polynomial calibration (Equation 4): for $\lambda(x) = ax^2 + bx + c$, we obtain $d\lambda/dx = 2ax + b$. This dispersion varies strongly with wavelength, ranging from $\approx0.051$\,nm/px at 706.5\,nm to $\approx0.292$\,nm/px at 447.1\,nm. The resolving power $R = \lambda/\Delta\lambda$, where $\Delta\lambda = \text{FWHM}_{\text{px}} \times (d\lambda/dx)$, declines from $R\approx142$ (red) to $R\approx16$ (blue). These values depend on grating density, pixel scale, and slit width. Consequently, narrow lines blend in the blue while broad continuum features remain well preserved. Accordingly, we bin the spectra to improve SNR (signal-to-noise ratio) for the objectives of this work.
\subsection{Observational Constraints}

All observations were conducted using the 11-inch telescope under clear sky conditions with moderate light pollution. Based on local meteorological parameters (high humidity, moderate wind speed) and typical boundary-layer turbulence for low-altitude sites in eastern India, the typical atmospheric seeing during observations is $\approx$1.5-3 arcsec, significantly larger than the diffraction limit of the 11-inch telescope. The spectrograph reliably records stellar spectra for targets brighter than V $\approx$ 3-4 mag. The limiting magnitude is primarily set by slit losses, spectral dispersion across the detector, and detector noise rather than photon collection alone. 

\subsection{Data Processing Pipeline}
We developed a comprehensive Python-based analysis pipeline to systematically process raw spectral images and extract calibrated stellar spectra. The pipeline integrates image preprocessing, wavelength calibration, spectral alignment, and stacking procedures.

\subsubsection*{Image Preprocessing}
Raw FITS (Flexible Image Transport System) images from the CMOS sensor were preprocessed to isolate the spectral signal from detector noise. Dark frame subtraction was performed using master dark frames acquired at matching exposure times and sensor temperatures to remove thermal noise and fixed-pattern detector artifacts. Traditional flatfielding was not applied because the dispersed spectrum occupies only a narrow horizontal band across the sensor, and pixel-to-pixel sensitivity variations are effectively normalized during the Vega-based flux calibration described below. 

Sky subtraction was not performed in this work. We acknowledge that atmospheric sky emission and absorption are present everywhere on the detector, and the recorded spectra therefore contain the combined signal from the source and the sky. Professional astronomical spectroscopy routinely includes sky subtraction to remove this contamination, typically by subtracting off-source sky frames or nodding between on-source and off-source positions. In our case, the decision to omit sky subtraction was driven by practical constraints in an educational/undergraduate research context, including limited observing time, telescope scheduling, and the focus on demonstrating fundamental spectroscopic techniques rather than publication-grade data reduction. This limitation has important consequences: some spectral features visible in our data, particularly narrow absorption lines not present in reference spectra, may represent telluric (atmospheric) contamination rather than intrinsic stellar features. All telluric features identified in this work are cross-referenced with the HITRAN2020 atmospheric line database \citep{HITRAN2020} and marked with green dashed lines in the spectral figures to distinguish them from intrinsic stellar absorption. We explicitly identify such cases in our analysis (Section 4) and caution readers that ground-based spectra without sky subtraction require careful interpretation when comparing with space-based observations.

To extract the one-dimensional spectrum, we summed pixel intensities along each row (perpendicular to dispersion direction) and fitted a Gaussian profile to identify the spectral trace center and width. The central region was extracted using a window defined by the full-width-at-half-maximum (FWHM) of the Gaussian fit, effectively rejecting pixels outside the primary spectral trace that may be contaminated by hot pixels or scattered light.

\subsubsection*{Correction for Instrument Sensitivity using Vega}
Vega ($\alpha$ Lyrae, spectral type A0V) served as our primary spectrophotometric standard for instrument sensitivity correction. We adopted the HST absolute spectrophotometry of Vega from \citet{bohlin2014} as our reference spectrum. This choice addresses the lack of manufacturer-provided sensor response curves by empirically calibrating instrumental sensitivity through comparison with a well-characterized stellar standard.

Prominent hydrogen Balmer absorption lines (H-$\alpha$ at 6563 \AA, H-$\beta$ at 4861 \AA) were identified in raw Vega observations and used as wavelength fiducials for initial wavelength mapping. To construct a wavelength-dependent sensitivity correction function $w(\lambda)$, we related observed intensities $I_{\text{obs}}$ to model intensities $I_{\text{model}}$ via:
\begin{equation}
w(\lambda) = \frac{I_{\text{model}}(\lambda)}{I_{\text{obs}}(\lambda)}
\end{equation}

The sensitivity correction function was smoothed using Savitzky-Golay filtering and interpolated across the full wavelength range to enable relative flux correction of subsequent stellar observations. However, since our stellar targets were observed on different nights and at different times under varying atmospheric conditions (air mass, seeing, transparency), and without sky subtraction or systematic air mass correction, this calibration corrects only for wavelength-dependent instrument sensitivity. It does not account for true absolute flux under varying observational conditions. Consequently, physically meaningful flux units (e.g., erg~s$^{-1}$~cm$^{-2}$~\AA$^{-1}$) cannot be reliably assigned to spectra of stars observed under different conditions than Vega. All spectra in this work are therefore presented in \textit{procedure-defined units} (p.d.u.), normalized to unity at their respective continuum peaks. This convention facilitates relative comparisons of spectral features and continuum shapes while acknowledging the limitations of single-standard calibration without atmospheric corrections. 

\subsubsection*{Wavelength Calibration via Spectral Alignment and Image Stacking}

Atmospheric seeing, tracking errors, and instrumental flexure introduce sub-pixel shifts between sequential exposures. To optimally align spectra before stacking and refine wavelength calibration, we employed $\chi^2$ minimization across identified spectral markers (typically H-$\alpha$ and H-$\beta$ absorption lines):
\begin{equation}
\chi^2 = \sum_{i=1}^{N} \sum_{j=1}^{N'} \frac{(\lambda_{i,j,\text{obs}} - \lambda_{j,\text{expected}})^2}{\lambda_{j,\text{expected}}}
\end{equation}
where $N$ is the number of images, $N'$ is the number of spectral markers, and shift parameters are optimized for each image to minimize total $\chi^2$.

\begin{figure}[H]
\centering
\includegraphics[width=\columnwidth]{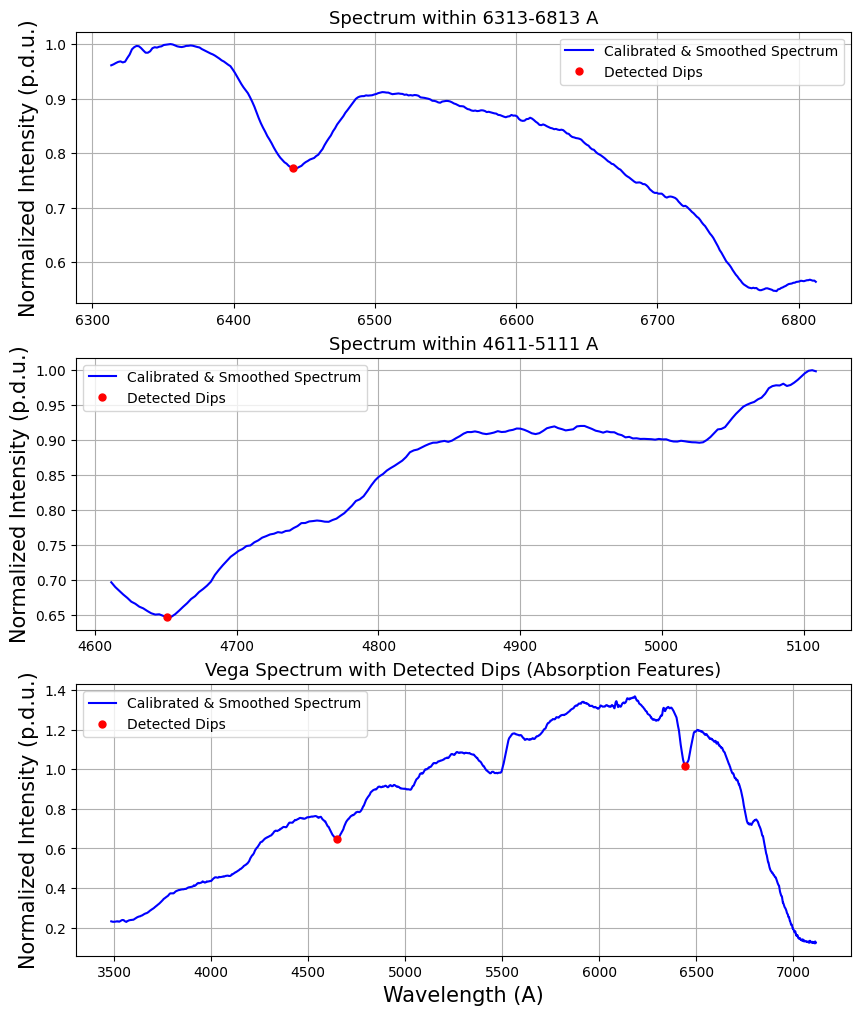}
\caption{Identification of H-$\alpha$ and H-$\beta$ spectral markers in a single frame of Vega observations. The algorithm identifies absorption line minima across multiple frames, enabling precise spectral alignment via $\chi^2$ minimization before stacking.}
\label{fig:spectral-markers}
\end{figure}

After alignment, spectra were binned into 1 Å wavelength intervals. For each bin, we calculated the weighted mean intensity and standard deviation across all images, yielding a high signal-to-noise final spectrum with associated uncertainty estimates - a feature uncommon in amateur spectroscopy but essential for quantitative analysis. Figure~\ref{fig:spectral-markers} illustrates the spectral marker identification process used for alignment.

\begin{figure}[H]
\centering
\includegraphics[width=\columnwidth]{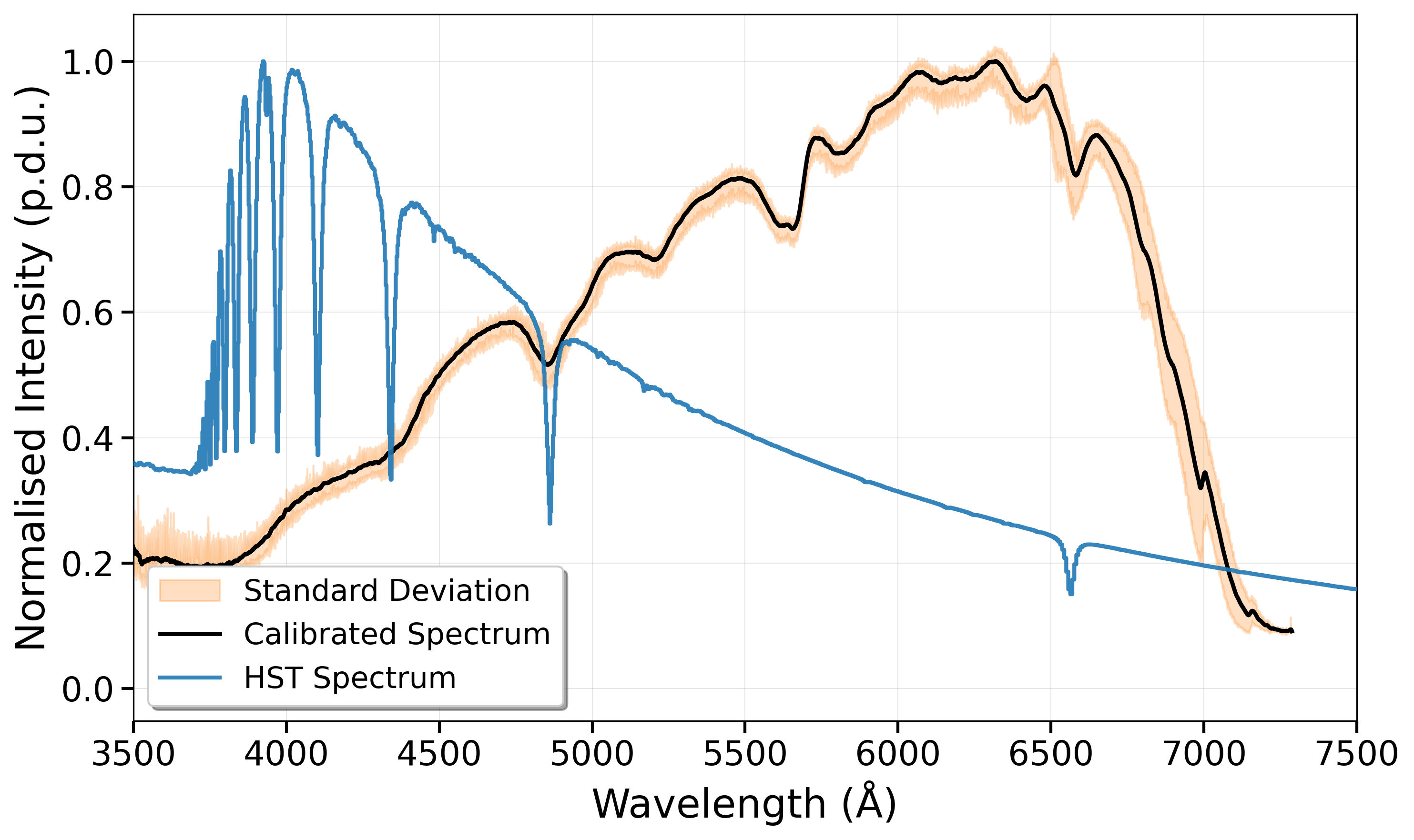}
\caption{Recorded Vega spectrum (black) after wavelength calibration compared with HST CALSPEC reference spectrum (blue) from \citet{bohlin2014}. The orange error bands in the recorded spectra represent standard deviations from stacked images, demonstrating the robustness of our data reduction pipeline. Please note that this spectrum has not yet been corrected for instrument sensitivity.}
\label{fig:vega-calibration}
\end{figure}


\section{Results}

All stellar spectra presented in this section were obtained using the 11-inch telescope (Table~\ref{equipment-table}). Typical individual exposure times ranged from 0.5 to 1 s, with 100 best frames stacked per target, yielding total integration times of approximately 100-200 s. 

\subsection{Spectral Features by Stellar Type}

Our spectrometer successfully captured calibrated spectra from five bright stars spanning spectral classes A through M (Table~\ref{stellarTargets}), with wavelength coverage from approximately 3500 to 7000 \AA.

\begin{table}[H]
\centering
\begin{tabular}{@{}ccc}
\hline
\textbf{Star} & 	\textbf{Spectral Class} & 	\textbf{Key Features} \\
\hline
Vega & A0V & Strong H lines \\
Sirius & A1V & Strong H lines \\
Procyon & F5IV-V & Metallic lines emerging \\
Capella & G8III + G0III & Strong metallic lines \\
Betelgeuse & M2Ia & TiO bands, Na doublet \\
\hline
\end{tabular}
\caption{Stellar targets observed with our spectrometer, covering spectral classes A through M}
\label{stellarTargets}
\end{table}

Procyon, Sirius, and Capella are known multiple-star systems as Procyon and Sirius are binary systems while Capella is a quadruple star system dominated by a spectroscopic binary. Given the slit width, seeing-limited spatial resolution, and f/10 focal ratio of the 11-inch telescope, the instrument does not spatially resolve individual components. The recorded spectra therefore represent composite light from unresolved stellar components, which may influence inferred continuum slopes and effective temperatures, particularly for systems such as Capella.

\begin{figure*}[p]
\centering
\includegraphics[width=0.75\textwidth]{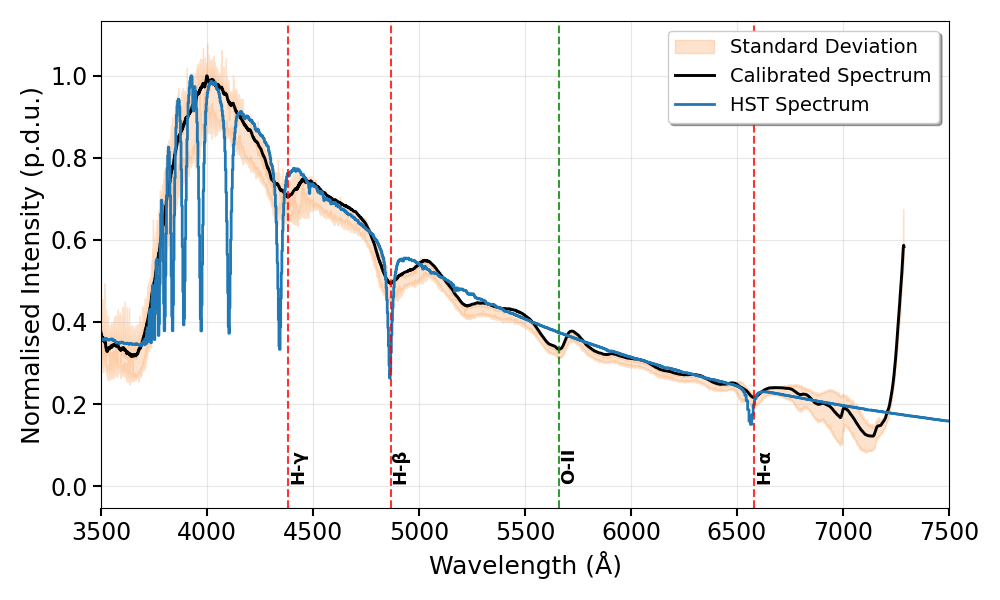}
\caption{Vega (A0V) spectrum (black) compared with HST reference spectrum (blue) from \citet{bohlin2014}. The spectrum displays prominent Balmer series absorption (H-$\alpha$, H-$\beta$, H-$\gamma$) characteristic of A-type stars. Major absorption features show excellent agreement with HST data, validating our calibration approach. The shaded region indicates the standard deviation from stacked frames. Flux is in p.d.u., normalized to the continuum peak.}
\label{fig:vega-spectrum}
\end{figure*}

\begin{figure*}[p]
\centering
\includegraphics[width=0.75\textwidth]{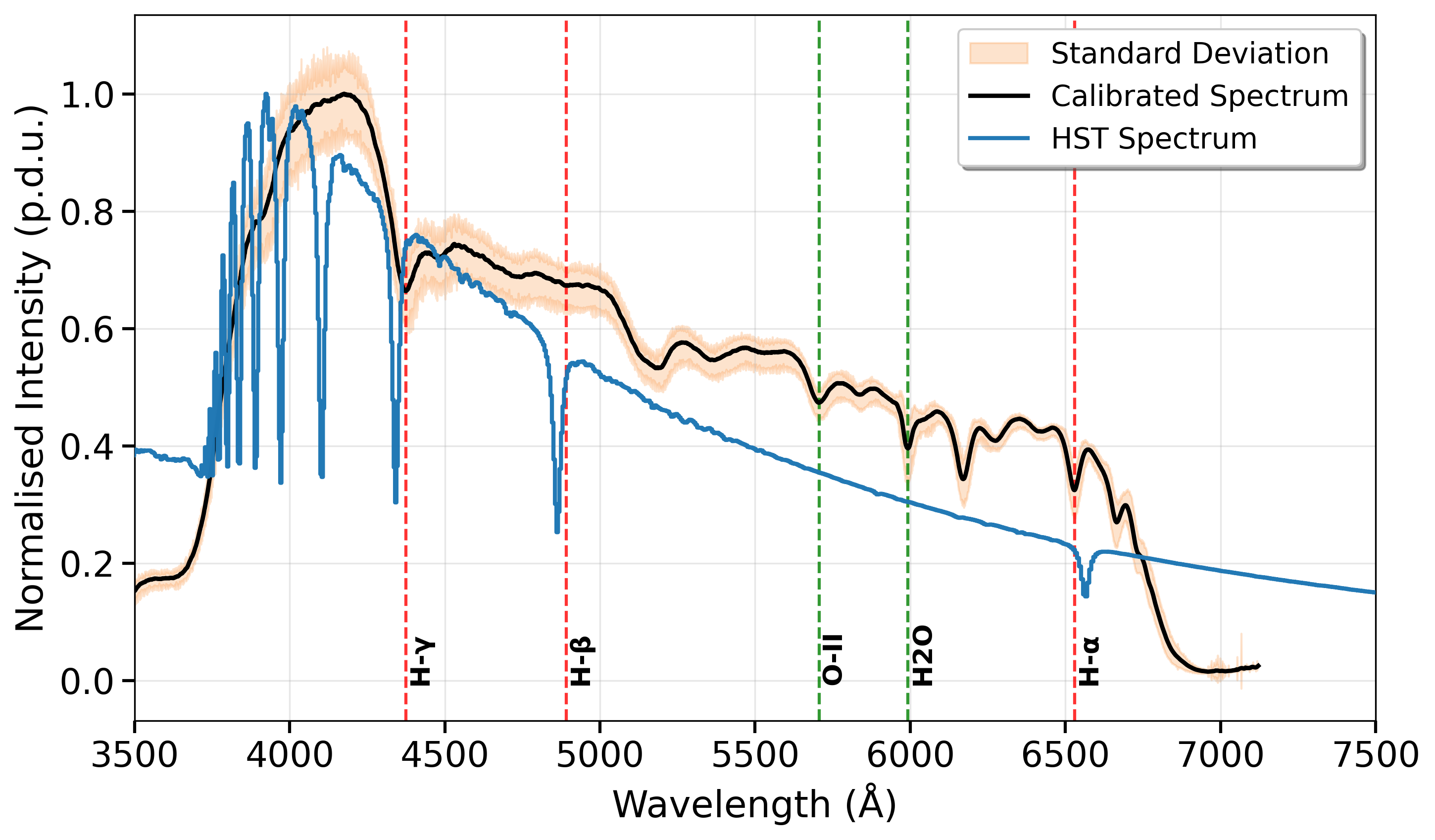}
\caption{Sirius (A1V) spectrum (black) compared with HST reference spectrum (blue) from the HST database \citep{bohlin2014}. The spectrum shows characteristic A-type stellar properties with prominent Balmer series absorption. Major absorption features align well with HST data. The shaded region indicates the standard deviation from stacked frames. Flux is in p.d.u., normalized to the continuum peak.}
\label{fig:sirius-spectrum}
\end{figure*}

\begin{figure*}[p]
\centering
\includegraphics[width=0.75\textwidth]{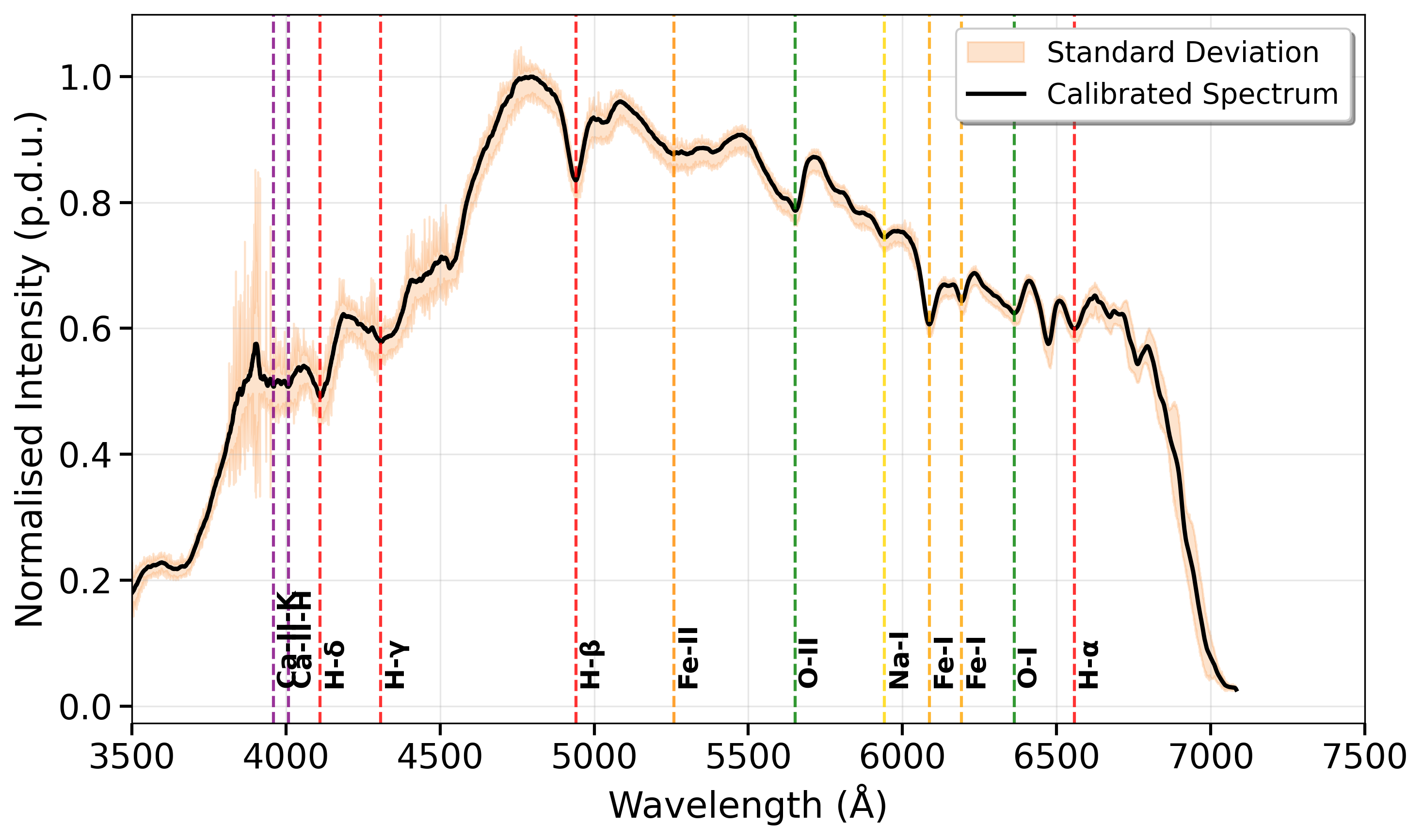}
\caption{Procyon (F5IV-V) spectrum showing weakened hydrogen lines relative to A-type stars and emerging metallic absorption features. Enhanced Ca~II H and K lines and multiple Fe~I features are visible. Flux is in p.d.u., normalized to the continuum peak.}
\label{fig:procyon-spectrum}
\end{figure*}

\begin{figure*}[p]
\centering
\includegraphics[width=0.75\textwidth]{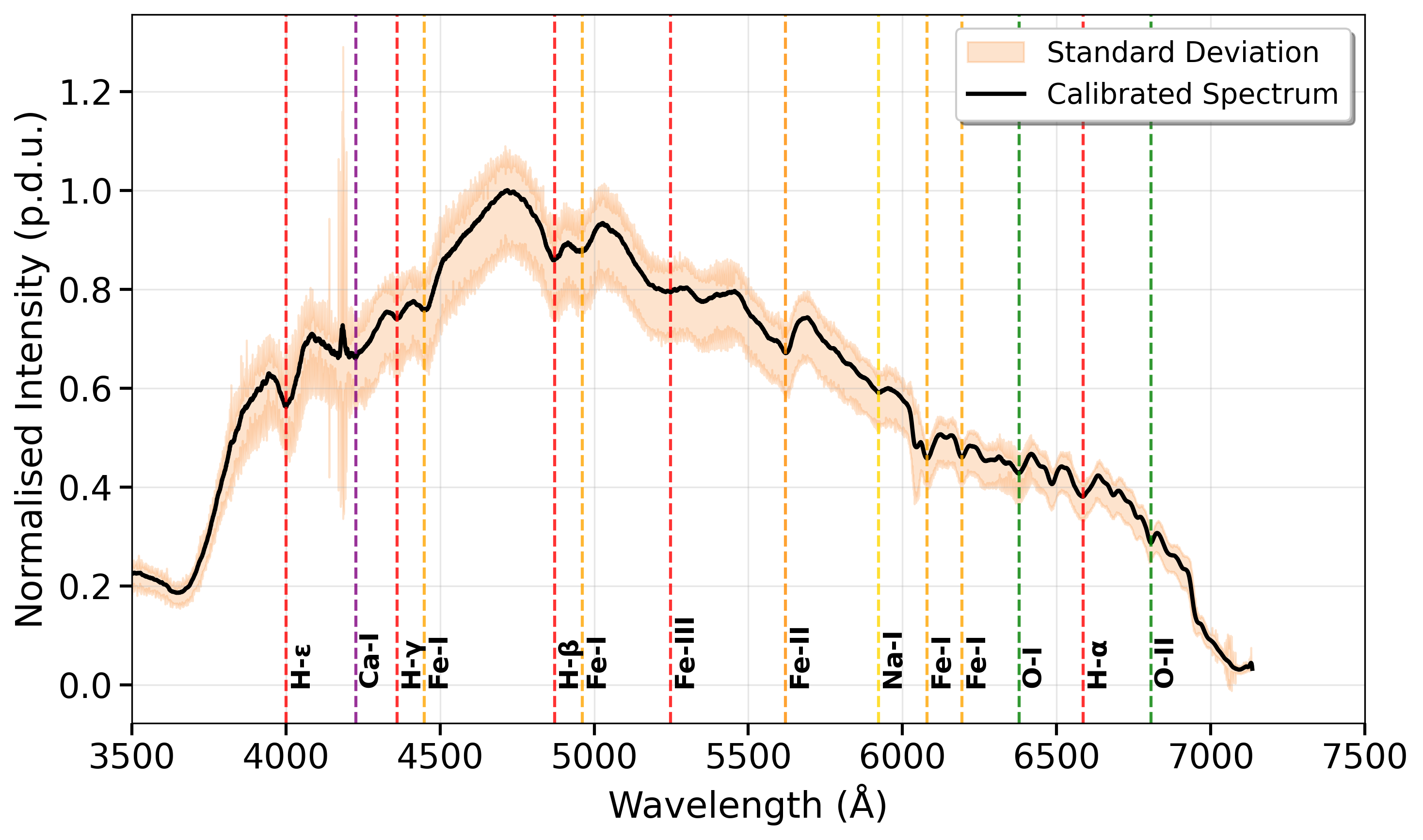}
\caption{Capella (G8III + G0III) spectrum dominated by metallic absorption lines. Multiple Fe~I and Fe~II lines are prominent in the blue-green spectral region along with the strong Ca~I line. Flux is in p.d.u., normalized to the continuum peak.}
\label{fig:capella-spectrum}
\end{figure*}

\begin{figure*}[p]
\centering
\includegraphics[width=0.75\textwidth]{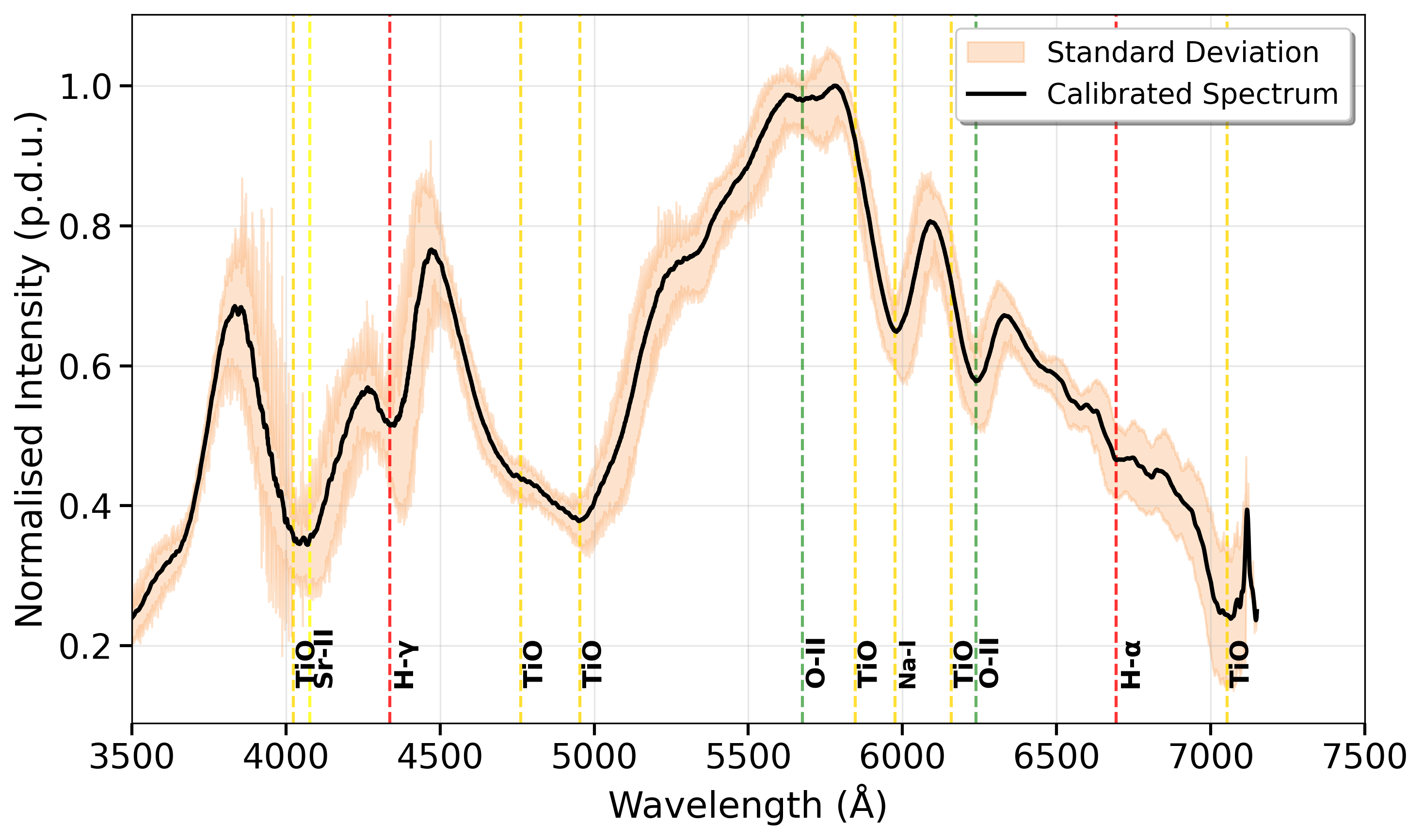}
\caption{Betelgeuse (M2Ia) spectrum showing weakened hydrogen absorption and strong molecular features, with prominent Na~I doublet and broad TiO absorption bands characteristic of cool M-type supergiant atmospheres. Flux is in p.d.u., normalized to the continuum peak.}
\label{fig:betelgeuse-spectrum}
\end{figure*}

\hfill

\subsubsection*{A-Type Stars (Vega and Sirius)}

Vega and Sirius (Figures~\ref{fig:vega-spectrum} and \ref{fig:sirius-spectrum}) exhibited the expected strong hydrogen Balmer absorption lines characteristic of A-type stellar atmospheres. We identified H-$\gamma$ (4340 \AA), H-$\beta$ (4861 \AA), and H-$\alpha$ (6563 \AA) as prominent features. Several additional absorption features are visible in Figures~\ref{fig:vega-spectrum} and~\ref{fig:sirius-spectrum} that are not present in the HST reference spectra and are not expected in A0V and A1V stellar atmospheres. These features, labeled as O-II (in both spectra) and H$_2$O (in Sirius), are identified as telluric atmospheric contamination, with H$_2$O representing water vapor absorption. These discrepancies directly reflect the limitations imposed by the absence of dedicated sky subtraction in the current observations.

\subsubsection*{F-Type Star (Procyon)}
Procyon (F5IV-V) (Figure~\ref{fig:procyon-spectrum}) - from a binary system with a faint white dwarf companion - showed weakening hydrogen lines relative to A-type stars, with H-$\beta$ depth reduced to $\sim$10\% and faint H-$\delta$ absorption at 4101 \AA. Metallic absorption lines became more prominent, including Ca-II H (3968\AA) and K (3933\AA) resonance lines, Fe-I lines (multiple features at 4450--6200 \AA), and enhanced Na I doublet (5890-5896 \AA)(Figure~\ref{fig:procyon-spectrum}). This trend toward metal-line dominance at lower temperatures aligns with Harvard classification criteria. Telluric atmospheric absorption features, primarily from oxygen (O-I \& O-II), are also present.

\subsubsection*{G-Type Star (Capella)}

Capella (G8III + G0III) (Figure~\ref{fig:capella-spectrum}) exhibited further weakening of hydrogen features and strengthening of metallic lines. A faint H-$\epsilon$ ($\sim$3970 \AA) can be seen. We can observe a prominent Ca-I feature ($\sim$ 4227 \AA) as expected with extended atmospheres. Multiple Fe I (neutral iron) and Fe II (singly ionized iron) lines dominated the blue-green spectral region (4000-6500 Å, Figure~\ref{fig:capella-spectrum}). Capella is an unresolved spectroscopic binary consisting of two giant stars (G8III and G0III).The observed spectrum therefore represents a flux-weighted superposition of the unresolved components Capella Aa and Ab, with effective temperatures of approximately 4970 K and 5730 K, respectively \citep{Torres_2015_Capella_bin}. Direct application of Wien’s displacement law to the composite spectrum is limited by the restricted wavelength coverage and the binary nature of Capella. However, approximate constraints can be obtained using a flux-weighted effective temperature,
\begin{equation}
    T_{fw} = \frac{\sum_i F_i T_i}{\sum_iF_i} 
\end{equation}

where $F_i$ denotes the bolometric flux contribution of the $i_{th}$ stellar component, yielding $T_{fw} \approx $ 5330K and hence $\lambda_{peak} \approx $ 5440\AA. This lies within the theoretical bounds of 5060-5830\AA  derived from the individual stellar components and agrees well with the observed spectral distribution. Now, the relative strengths of hydrogen and metallic absorption features provide a robust basis for classifying the composite spectrum as G-type. As with the other spectra, telluric contamination from atmospheric oxygen is also present.

\subsubsection*{M-Type Star (Betelgeuse)}

Betelgeuse (M2Ia) (Figure~\ref{fig:betelgeuse-spectrum}) displayed the coolest spectrum with weaker hydrogen absorption and strong molecular absorption features. The Na-I doublet appeared prominently in emission or partial absorption-filling, characteristic of extended supergiant atmospheres, which also favors singly ionized strontium (Sr-II) feature ($\sim$ 4078 \AA) with low excitation energy,(Figure~\ref{fig:betelgeuse-spectrum}). While our spectral resolution was insufficient to fully resolve all the TiO (titanium oxide) molecular lines \citep{betel_tio_bands}, we observed broad absorption features near 4800 \AA\ and 6000 \AA consistent with the band structure of TiO.
For Betelgeuse ($\sim$3800 K)\citep{BGtem_stac1969}, Wien's displacement law predicts a spectral peak near 7626 \AA, which is beyond the wavelength range of this spectrograph. Since the observed spectrum is limited to the wavelength range $\approx$ 3500-7000 \AA, the full spectral energy distribution (SED) of the system is not sampled. Telluric absorption features from $O_2$ are also present, distinguishing atmospheric contamination from the intrinsic molecular bands and atomic lines of this cool supergiant.


\subsection{Comparison with Observational Data}

Direct comparison of our Vega and Sirius spectra with HST observations revealed good correspondence in major absorption features despite the difference in aperture and detector sophistication (Figures~\ref{fig:vega-spectrum} and \ref{fig:sirius-spectrum}). We focus the HST comparison on these two targets because they were observed most extensively in our dataset, with longer cumulative exposure times (several hours total) that yielded higher signal-to-noise ratios suitable for detailed comparison. However, quantitative comparison via equivalent width measurements is performed only for Sirius. Since Vega was used as the sensitivity correction standard, including it in the quantitative comparison would introduce systematic bias as any residual calibration errors would be minimized in the Vega spectrum by construction. Additionally, the lack of sky subtraction in our data introduces telluric contamination, which manifests as spurious absorption features in ground-based spectra that are absent in space-based HST observations. This is particularly evident in the Sirius spectrum, where some narrow features not present in HST data originate from atmospheric absorption rather than the stellar photosphere.

To provide a resolution-independent quantitative comparison for Sirius, we measured equivalent widths (EW) for the Balmer lines H-$\alpha$, H-$\beta$, and H-$\gamma$ in our calibrated spectrum and compared them to values measured from HST reference data (Table~\ref{tab:ew-comparison}).

\begin{table}[H]
\centering
\begin{tabular}{ccccc}
\hline
\textbf{Line} & $\lambda_{\text{std}}$ (\AA) & \textbf{EW$_{\text{cal}}$ (\AA)} & \textbf{EW$_{\text{HST}}$ (\AA)} & \textbf{Ratio} \\
\hline
H-$\alpha$ & 6563 & 6.5 & 11.4 & 0.57 \\
H-$\beta$ & 4861 & 0.5 & 16.4 & 0.03 \\
H-$\gamma$ & 4341 & 8.9 & 16.2 & 0.55 \\
\hline
\end{tabular}
\caption{Equivalent width measurements for Sirius Balmer lines comparing our calibrated spectrum (EW$_{\text{cal}}$) with HST reference data (EW$_{\text{HST}}$). The ratio (EW$_{\text{cal}}$/EW$_{\text{HST}}$) indicates a systematic underestimation of the measured absorption line strength in our data.}
\label{tab:ew-comparison}
\end{table}

Our calibrated spectrum yields EW values systematically lower than HST measurements, with H-$\alpha$ and H-$\gamma$ at approximately 55--57\% of HST values. The H-$\beta$ measurement is severely underestimated (ratio $\approx$0.03). Examination of the raw spectral data revealed that this is not a wavelength calibration issue but rather a consequence of sensor saturation. Due to Sirius's extreme brightness, the continuum flux near the H-$\beta$ region (4861~\AA) saturates the CMOS sensor, effectively suppressing the line contrast and making the absorption feature nearly undetectable above the saturated continuum. This saturation effect is localized to the blue-green region where Sirius's spectral energy distribution peaks most strongly for an A1V star. While improved flux calibration techniques could potentially mitigate this effect, such refinements are currently beyond the scope of the current work. This highlights an important limitation when observing exceptionally bright targets with modest sensor dynamic range. Furthermore, the absence of sky subtraction means that telluric emission contributes to the measured continuum level, which inflates the apparent continuum and systematically reduces the measured EW of all absorption lines relative to sky-subtracted HST data. The systematic underestimation for H-$\alpha$ and H-$\gamma$ likely results from: (1) instrumental broadening spreading line flux beyond the integration window; (2) continuum placement challenges in lower-resolution data; and (3) incomplete flux calibration, particularly in the bluer regions where sensor response is typically weaker. The primary differences were decreased signal-to-noise in our data and stronger telluric contamination, both expected for small-aperture ground-based observations.

The dense forest of narrow absorption lines visible in HST spectra in the blue region (3500--4500~\AA) is largely absent in our data due to the limited resolving power ($R \approx 16$--$142$, wavelength dependent) of the instrument. At this resolution, closely spaced lines are blended into broader features, suppressing fine structure while preserving strong lines, continuum slopes, and molecular bands.

\section{Discussion}

Our stellar spectrometer successfully demonstrated the feasibility of spectroscopy using a low-cost spectrometer module paired with existing institutional telescope infrastructure.

\subsection{Calibration Performance and Accuracy}

The wavelength calibration accuracy is typically $\pm$2--3~nm ($\pm$20--30~\AA) across the visible range, with larger deviations toward the blue where dispersion per pixel is highest. The apparent absence or severe weakening of H-$\beta$ in the Sirius spectrum does not indicate a wavelength calibration error; examination of raw data revealed this is caused primarily by sensor saturation in the blue-green continuum region where Sirius is exceptionally bright. The saturated continuum suppresses the line contrast, making the absorption feature nearly undetectable. Additionally, the low resolving power ($R\approx20$--30) at 4861~\AA\ causes blending with adjacent spectral features, further degrading the detection and making continuum placement ambiguous. These combined effects represent a significant limitation when observing extremely bright targets with sensors of limited dynamic range and modest resolving power. As evident in the raw wavelength-calibrated Vega spectrum (Figure~\ref{fig:vega-calibration}), the CMOS sensor exhibits significantly higher sensitivity towards the red end of the spectrum than the blue, further suppressing blue-wavelength features and contributing to the apparent weakness of H-$\beta$ and H-$\gamma$ relative to H-$\alpha$. Similarly, the absence of Ca~II H and K resonance lines in the Capella spectrum, which are intrinsically strong in G-type giants, is attributed to a combination of declining instrumental sensitivity toward shorter wavelengths, atmospheric extinction, and residual sky contamination from the lack of sky subtraction. These lines remain identifiable in the F-type spectrum of Procyon, where the higher signal-to-noise ratio and different atmospheric conditions during observation preserved sufficient line contrast. 

\vfill

\subsection{Systematic Limitations}

Several absorption features visible in the ground-based spectra are absent in space-based HST observations, confirming their non-stellar origin. In particular, features labeled as O-II and H$_2$O in the Sirius spectrum (Figure~\ref{fig:sirius-spectrum}) are identified as telluric atmospheric contamination rather than intrinsic stellar features \citep{HITRAN2020}. Across all stellar spectra, telluric absorption bands arising primarily from H$_2$O and O$_2$, particularly in the 6500--6600~\AA\ region and near 6867--6884~\AA, are marked by green dashed lines \citep{HITRAN2020}. These features are consistently present across different stellar targets and are attributed to residual atmospheric absorption resulting from the lack of dedicated sky subtraction. The absence of sky subtraction is a principal source of discrepancy between our spectra and HST reference data, and some features labeled in the spectra of Procyon, Capella, and Betelgeuse (Figures~\ref{fig:procyon-spectrum}--\ref{fig:betelgeuse-spectrum}) may similarly include atmospheric rather than stellar contributions. Only robust spectral features that are consistently detected and expected for the given spectral type are interpreted as intrinsic stellar absorption lines.

These combined effects---the absence of sky subtraction introducing telluric contamination, sensor saturation for bright targets, blending at low resolution, wavelength-dependent sensitivity, and spectral complexity---explain the observed differences from HST data without invoking large systematic wavelength errors. The agreement between our Vega and Sirius observations with HST reference data for the stronger, well-isolated features validates both our calibration methodology and data reduction pipeline.

\subsection{Instrument Design Evaluation}

The choice of Vega as a spectrophotometric standard proved particularly effective, addressing the absence of manufacturer-provided sensor response curves through implicit calibration. This approach enabled quantitative comparison with established stellar standards while maintaining accessibility for educational applications. The $\chi^2$ minimization technique for spectral alignment, combined with 1 \AA\ binning and uncertainty propagation, produced results remarkably similar to HST observations, which is an excellent result for amateur spectroscopy and demonstrates the potential of accessible instrumentation.

Several design choices proved critical to instrument performance. The 600 lines/mm grating provided optimal dispersion for our small sensor, while first-order diffraction avoided spectral overlap complications. The precision slit system (1.0-1.5 mm) effectively balanced light throughput against spectral resolution, though narrower slits introduced problematic Fresnel diffraction artifacts. The 3D-printed mount's mechanical stability and telescope interface design enabled consistent results across multiple observing sessions.

The design approach is intentionally flexible and can be adapted to different equipment configurations. For the diffraction grating, we recommend mid-range line densities (500--750 lines/mm) as a starting point: lower densities sacrifice spectral resolution, while higher densities compress the wavelength coverage onto the sensor. Readers with access to larger sensors (e.g., $>$50~mm width) may consider second-order diffraction ($n=2$), which doubles the angular dispersion and improves resolution at the cost of reduced intensity and potential order overlap requiring appropriate filtering. Our use of first-order diffraction was dictated by the compact Skyris 236M sensor (5.44~mm width); a larger sensor would permit either extended wavelength coverage in first order or improved resolution in second order. Larger sensors also simplify optical alignment and relax the grating-to-sensor distance constraints, though at the expense of instrument compactness. Time and equipment constraints prevented us from exploring these alternatives, but they represent promising avenues for future implementations seeking higher resolving power.

\subsection{Pedagogical Implications and Recommendations}

The measured resolving power reveals important instrument characteristics. The FWHM remains approximately constant in pixel units (mean $\approx101\pm8$\,px), indicating that spectral broadening arises primarily from the slit/optical image size rather than wavelength-dependent effects. However, wavelength-dependent dispersion, ranging from $\approx0.051$\,nm/px ($0.51$\,\AA/px) at 706.5\,nm to $\approx0.292$\,nm/px ($2.92$\,\AA/px) at 447.1\,nm, causes resolving power to decline from $R\approx142$ in the far red to $R\approx16$ in the blue. Narrow lines are thus more blended in the blue, though broad continuum trends and molecular bands (evolving on scales of tens to hundreds of \AA) remain well preserved across 3500--7000\,\AA. For educational use, we recommend reporting wavelength-dependent $R(\lambda)$ values, presenting raw and binned data to illustrate the S/N versus resolution trade-off, and emphasizing the pixel-limited nature of the instrument. These considerations set realistic expectations: continuum fitting, synthetic color indices, and coarse classification remain appropriate, whereas line-profile analysis or radial-velocity work lie beyond the instrument's capabilities.

Key limitations include restricted wavelength coverage (3500-7500 \AA), preventing near-UV and near-IR observations, and insufficient resolution for fine spectral structure. The omission of sky subtraction introduced telluric contamination across all spectra, complicating the identification of weak stellar features and introducing spurious absorption lines of atmospheric origin. Sensor saturation for exceptionally bright targets such as Sirius further limited detection of features in the blue-green continuum region. The magnitude limit (V $<$ 3 mag) restricts targets to bright stars, though this aligns well with educational objectives. Future implementations should incorporate sky subtraction as a standard reduction step to enable cleaner spectral comparisons with reference data.

We have successfully constructed and characterized a compact, low-cost stellar spectrometer that produced reliable results across multiple spectral classes. Observations of five stars spanning spectral types A0V through M2Ia revealed the expected progression of spectral features: strong hydrogen absorption in A-type stars, the emergence of metallic lines in F-type stars, prominent Ca-II, H and Fe(I and II) features in G-type giants, and molecular absorption bands in M-type supergiants. The image-processing pipeline, which incorporated $\chi^2$-based spectral alignment and uncertainty estimation, yielded consistent and reproducible results. Overall, this work demonstrates the feasibility of conducting meaningful stellar spectroscopy with accessible equipment, emphasizing its potential as a practical tool for educational and undergraduate research in observational astronomy.

\section*{Acknowledgements}

We thank Dr. Kartik Senapati for his supervision, guidance, and encouragement throughout this project, and Dr. Gunda Santosh Babu for valuable feedback on observational procedures. We acknowledge the technical support from Mr. Rudranarayan Mohanty and Mr. Sakthivel V. A. for laboratory resources and equipment access. We also gratefully acknowledge the Integrated Lab (Physics) setting at School of Physical Sciences, NISER for providing the opportunity and resources to propose, develop, and undertake this undergraduate experimental project. This work would not have been possible without the generous support of the NISER Astronomy Club\footnote{NISER Astronomy Club \url{astroclub-niser.github.io}} (NAC), which provided access to the Skyris 236M CMOS sensor and both the 8-inch and 11-inch reflecting telescopes. We are particularly grateful to Vasanth Kashyap for assistance with spectrometer design and 3D modeling, and Abha Vishwakarma for insightful discussions on data analysis methods. The NISER RoboTech Club\footnote{NISER Robotech Club \url{rtcniser.github.io}} manufactured the spectrometer mount via 3D printing. We thank the NAC members for assistance with telescope setup during nighttime observing sessions.

\section*{Data Availability}

The raw spectral data and the processed spectral datasets obtained using the telescope for stellar observations and calibration sources are available from the authors upon reasonable request. The data processing pipeline and analysis code used in this work are publicly available in our GitHub repository at \url{https://github.com/niti21/Stellar-Spectrometry}.


\bibliography{ref}{}
\bibliographystyle{aasjournal}

\end{document}